\documentclass[journal]{IEEEtran}
\usepackage{amsmath,amssymb,amsthm,amsfonts}
\usepackage{graphicx}
\usepackage{xcolor}
\usepackage{epstopdf}
\usepackage{algorithm}
\usepackage{algpseudocode}
\usepackage{mathtools}
\usepackage{enumitem}
\usepackage{caption}

\usepackage{wrapfig}
\usepackage{hyphenat}
\usepackage{flushend}

\newtheorem{thm}{Theorem}

\newtheorem{cor}{Corollary}

\newtheorem{remark}{Remark}

\newcommand{\vv}[1]{\mathbf{#1}}

\newcommand{\E}[1]{E \left \{#1 \right\}}

\newcommand*{\tran}{^{\mkern-1.5mu\text{T}}}

\DeclarePairedDelimiter\floor{\lfloor}{\rfloor}
\newcommand{\titi}[1]{\textcolor{black}{#1}}
\newcommand{\tete}[1]{\textcolor{black}{#1}}

\usepackage{ifpdf}
\ifpdf
\usepackage{epstopdf}
\fi

\begin{document}

%
%
%

\title{Bit Density Based Signal and Jamming Detection in 1-Bit Quantized MIMO Systems}
\author{M.~A.~Teeti
	\thanks{M. Teeti is with School of Information Engineering, East China University of Technology, Nanchang, 330013, China (e-mail: 201960077@ecut.edu.cn)}
}%
\maketitle

\begin{abstract}
This paper studies the problem of deciding on the absence (i.e., null hypothesis, $\mathcal{H}_0$) or presence (i.e., alternative hypothesis, $\mathcal{H}_1$) of an unknown signal embedded in the received signal in a multiple-input, multiple-output (MIMO) receiver, employing 1-bit quantization. The originality of our solution lies in quantizing the received signal by an adapted 1-bit window comparator, rather than a traditional 1-bit quantizer. This enables us to divide the space of observed binary sequences into two typical sets (w.r.t. the distribution of the no. of 1’s in a sequence) asymptotically, where the first set corresponds to $\mathcal{H}_0$ and the second to $\mathcal{H}_1$. As a result, we reduce the detection problem to determining the highly probable set for an observed sequence. Thus, a very low-complexity binary hypothesis detector is proposed and its probability of detection is given. To show the high efficacy of the proposed 1-bit receiver structure, we consider two wireless applications; jamming detection in a massive MIMO system, and probing a non-stationary low-power transmitter in a wireless sensor network (WSN), assuming unknown Rayleigh-fading channels. Compared with an unquantized system employing a chi-square test, it is shown that the performance loss can be roughly as large as 10\% in massive MIMO and this gap diminishes as sequence length or/and jamming power increases. For WSN, we show that compared with an unquantized system, the performance gap becomes smaller when the observation interval is extended over a few symbols.
\end{abstract}



\section{Introduction}
\label{sec:section1}

The need for efficient and reliable mechanisms for securing multiple-input, multiple-output (MIMO) systems has been brought on in recent years by the increased demand for safeguarding massive MIMO systems against malicious attacks \cite{Zhu2014, 7108032, 7491252, Kapetanovic2015, Goel2008}. Many attempts have shown that passive attack (eavesdropping) in massive MIMO is in principle avoidable asymptotically \cite{Kapetanovic2015}, whereas active attack (jamming attack) is more deleterious, especially pilot jamming. Therefore, jamming attack is a crucial challenge that incurs a significant loss in performance if left undetected.

In \cite{Julia2016, julia2017}, some asymptotic results in random matrix theory are leveraged to detect and suppress jamming under the assumption that signal and jamming reside in a very small-dimensional subspace compared to the number of BS antennas. The authors in \cite{julia2017} showed that when the dimension of jamming subspace increases, jamming detection turns to be difficult (if not impossible). In \cite{7010928}, the asymmetry of received signal power levels between the legitimate parties is exploited to develop energy-based detector for detecting pilot attack. An attempt to use pilot hopping and an increased size of pilot set in the uplink of massive MIMO system has been considered in \cite{Yuksel2015} to mitigate the impact of pilot attack. In \cite{Wu2018}, pilot-data based channel acquisition is considered mitigating pilot attack in the asymptotic sense of data length. Pilot re-transmission and adaptation techniques are considered in \cite{7750607} to tackle pilot attack.

In  \cite{XU2020107411,Akhlaghpasand2018}, jamming detection is formulated as a binary hypothesis testing, and hence pilot attack detection has been studied under different conditions. Reference \cite{XU2020107411} uses different approaches of hypothesis testing to detect a pilot spoofing originating from a single-antenna jammer in a massive MIMO system. The authors in \cite{Akhlaghpasand2018} consider a generalized likelihood ratio test (GLRT) over many coherence blocks of channel while taking advantage of unused pilots to detect a multiple-antenna jammer. However, there is a tradeoff between detection capability and delay. The estimation of a single-antenna jammer’s statistic is considered in \cite{Akhlaghpasand2020d} using unused pilots in the system for jamming suppression. The extension to a multiple-antenna jammer is still lacking. A linear channel estimator and bilinear equalizer in data phase for correlated massive MIMO is studied in \cite{Akhlaghpasand2020c} to improve performance in the presence of a single-antenna jammer.

1-bit quantized MIMO systems where antennas (collocated or distributed) are equipped with 1-bit analog-to-digital converters (ADCs) instead of high resolution ADCs has been gaining importance in recent years \cite{7586074, 6804238, 7094595, T2018, shao2017, shao2020} because of the remarkable simplification of physical layer and reduction in energy consumption \cite{7000981}. The interest in signal quantization has also been an important research topic in wireless sensor networks (WSN) \cite{6422327,Aalo1994MultilevelQA}, motivated by energy and bandwidth constraints in such networks \cite{Howard2006}. In WSN, a relevant problem to this work is that of detecting the presence or absence of a low-power transmitter (declaring its presence periodically) by a set of cheap sensors, where these sensors quantize their observations to a single bit before sending them to a fusion center for processing.

It was shown in \cite{Teeti2020} that pilot attack can have a deleterious impact on performance of massive MIMO using 1-bit ADCs. Since the polarity ($\pm 1$) of the signal is only available at the BS, jamming detection appears to be much intricate when compared with the unquantized system. To the best of the author's knowledge,  studies on jamming detection in 1-bit quantized MIMO systems are still lacking. Besides the high non-linearity of traditional 1-bit quantizer, a stronger jammer can quickly saturate its output. Unfortunately, the existing jamming detection approaches developed in the literature for unquantized systems seem inapplicable under 1-bit quantization.

The aim of this work is to  provide a binary quantization method in a MIMO receiver which facilitates binary hypothesis testing, and also develop a low-complexity detection method. The proposed quantization and binary hypothesis detection methods will bring many advantages, features and new different applications, including jamming detection in 1-bit quantized massive MIMO system and the detecting the presence of a low-power transmitter in WSN. In such systems, 1-bit quantization is a very attractive solution for its low-hardware complexity and low-energy consumption. Also, it is hoped that this work will stimulate further research on the subject. 

We summarize the major contributions of this work as follows.

\begin{enumerate}[leftmargin=*]
	\item  We propose an adapted 1-bit window comparator for quantizing the incoming signal, giving rise to a Bernoulli process in the \emph{null hypothesis} ($\mathcal{H}_0$) with fixed success rate and another Bernoulli process in the \emph{alternative hypothesis}($\mathcal{H}_1$) with a monotonically increasing success rate with an increasing variance of the model for $\mathcal{H}_1$. 
	
	\item  A very low-complexity algorithm called \emph{ bit density detection} (BDD) algorithm is developed where the binary hypothesis detection is basically based on counting the number of 1’s in an observed binary sequence. Also, the detection probability is derived. Further, only two parameters are fed to the BDD algorithm, where these parameters are computed offline.   
	
	\item  The proposed 1-bit window comparator combined with the BDD algorithm is leveraged for jamming detection in a 1-bit quantized massive MIMO system and for probing a low-power transmitter in a WSN. The results in both cases show that the proposed detection mechanism is promising.
\end{enumerate}

\tete{We stress that throughout this paper; we assume that the variance of the model in $\mathcal{H}_0$ is known at the receiver, whereas the variance of the model in $\mathcal{H}_1$ is assumed unknown.}

The rest of this paper is organized as follows. Sec. \ref{sec:section2} \tete{provides some motivation for this work} and presents the mathematical formulation of the problem. Sec. \ref{sec:section3}  presents the proposed 1-bit receiver structure, comprising 1-bit window comparator and binary hypothesis detector, and the major results are given. In Sec. \ref{sec:section4} we utilize the proposed 1-bit receiver structure for jamming detection in massive MIMO system and for signal probing in WSN. Sec. \ref{sec:section5} shows some numerical results to verify the calculations, and Sec. \ref{sec:section6} summarizes this paper.

\section{Motivation and Problem Formulation}
\label{sec:section2}

\subsection{Motivation}

In the literature of MIMO communication, a traditional zero-threshold 1-bit quantizer (aka. 1-bit ADC or comparator) has been largely investigated. In typical wireless communication, Rayleigh fading, the randomness of data, interference, and background noise give rise to a received signal admitting a symmetrical or slightly skewed density around zero. If a traditional 1-bit quantizer is used to quantize the received signal, then the number of observed 1’s and 0’s averaged over all randomness is roughly equal. \tete{It is even more interesting to note that} a balance between 1’s and 0’s is expected over a single realization of the channel when the receiver is equipped with a sufficiently large number of antennas, thanks to the law of large numbers.

\tete{Binary hypothesis testing in communication systems is one of the most important problems. With binary observations,} a receiver needs to decide which one of two hypothetical models gave rise to the observed binary sequence, i.e., detecting the presence or absence of an unknown signal (e.g., undesired jamming signal) in the received observations. \tete{In the light of the presence of such undesired signal and coarsely quantized observations, it seems natural to assume that the channel is unknown, since channel estimate may no longer be useful under this situation. With the above discussion in mind, it turns out that,} with a traditional 1-bit quantizer, extracting information about a presence of an unknown signal (e.g., jamming) from the observed digits seems intricate. Further, it is unknown yet that using higher-order statistics can lead to a simple detection technique due to the inherent high non-linearity of the traditional quantizer. Besides, a stronger undesired signal such as malicious jamming can quickly saturate its output. Therefore, our proposed binary hypothesis detection mechanism in the next section brings many advantages, comprising low-complexity, simplicity, and usefulness in different applications.

Let $b_1,b_2,\cdots, b_n$ be a binary sequence with independent and identically distributed (i.i.d.) random variables. It is easy to think of this sequence as if it was generated by one of two discrete memoryless sources ${S}_0$ and $ {S}_1$. Let the \emph{null} ($\mathcal{H}_0$) and \emph{alternative} ($\mathcal{H}_1$) hypotheses correspond to a binary sequence $\{b_i\}$ generated by the source ${S}_0$ and ${S}_1$, respectively. Further, let $\theta_0 \triangleq \mathbb{P}(b_i = 1|\mathcal{S}_0)=0.5$ and $\theta_1\triangleq \mathbb{P}(b_i=1|\mathcal{S}_1)=0.5$ be the success rates of the corresponding binary sources. Thus, both sources have the same \emph{entropy}, i.e., $H_b(\theta_0) = H_b(\theta_1) = 1$, \tete{where $H_b(\theta) =-\theta \log_2(\theta)-(1-\theta)\log_2(1-\theta)$ denotes the binary entropy function of a Bernoulli random variable with success rate $\theta$}. For a sufficiently large $n$, the \emph{typical sets} (\tete{i.e., sets of high probability sequences}) $\mathcal{T}_0^n$ and  $\mathcal{T}_1^n$ of sequences generated by ${S}_0$ and ${S}_1$ admit the same cardinality asymptotically \footnote{\tete{For an i.i.d. Bernoulli process of success rate $\theta$, the cardinality (size) of a typical set of binary sequences each of length $n$ generated by the process is roughly $2^{n H_b(\theta)}$ \cite{ElementsofIT}}.}, i.e., $|\mathcal{T}_0^n|=|\mathcal{T}_1^n|\approx 2^n$, \tete{where $|A|$ denotes the cardinality of a set $A$}. Since both typical sets span the whole space of binary sequences (i.e., the two typical sets are almost fully overlapping), there will be no hope to divide the space of binary sequences into two disjoint regions from which one can distinguish if an observed binary sequence $\{b_i\}$ was likely generated under $\mathcal{H}_0$ or $\mathcal{H}_1$.

\tete{Analogous to the typical sets in information theory, our proposed quantization method enables us, in the asymptotic sense, to divide the set of all binary sequences at the output of 1-bit quantizer into two typical sets under both hypotheses (say, presence and absence of an unknown signal). In our case, we take a typical set with regard to the distribution of the number of 1’s in a sequence.}

Specifically, we propose to use an adapted 1-bit window comparator (aka window detector) to induce some imbalance between the number of $1$’s and $0$’s at its output under both hypotheses. For instance, the density of 1’s (i.e., the ratio between the number of 1’s and the total number of bits in the observed binary sequence) under $\mathcal{H}_0$ is kept fixed at some small value, say, $\theta_0$. \tete{Conditioned on $\mathcal{H}_0$, the number of 1’s in an observed sequence converges to $n\theta_0$ (in probability), thanks to the weak law of large numbers. This means that the null typical set comprises all binary sequences, where the number of 1's in each sequence is roughly $n\theta_0$. Hence, for an observed sequence with the number of 1’s which deviates from $n \theta_0$ will be highly likely caused by a model in $\mathcal{H}_1$, i.e., the observed sequence will belong to the alternative typical set.  }

\begin{figure}[!ht]
	\centering
	\includegraphics[scale=0.8]{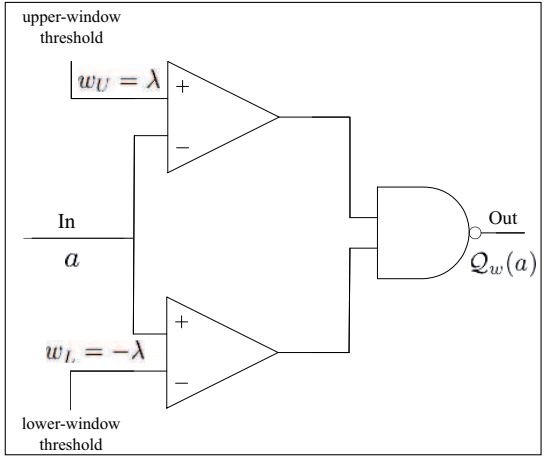}
	\caption{Logic-gate diagram of a window comparator. }
	\label{fig:WC} 
\end{figure}

\begin{figure}[!ht]
	\centering
	\includegraphics[scale=0.8]{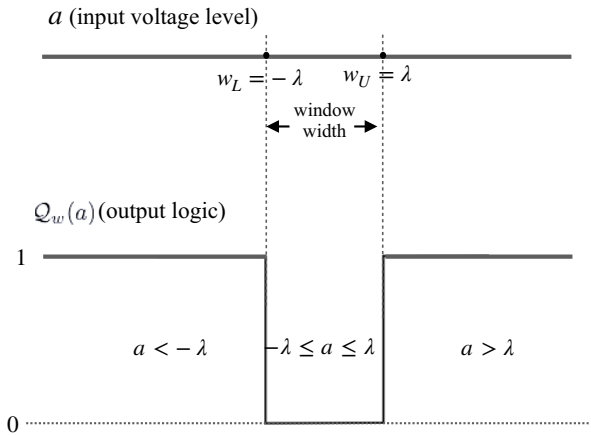}
	\caption{The operation of the window  comparator in Fig. \ref{fig:WC}.}
	\label{fig:WC_operation} 
\end{figure}

 \subsection{Problem Formulation}
 
Fig. \ref{fig:WC} shows a logic gate diagram of our memoryless 1-bit window comparator, comprising two operational amplifiers working as comparators and a NAND gate. The upper- and lower-window thresholds are given by $w_U=\lambda$ and $w_L=-\lambda$, respectively. With reference to Fig. \ref{fig:WC_operation}, the main operation of the window comparator in Fig. \ref{fig:WC} is to detect whether an input voltage is outside a window predetermined by the upper- and lower-window thresholds. This can be mathematically defined by
 \begin{equation}
 	\label{eq:window_detector}
 	\mathcal{Q}_w (a)=
 	\begin{cases}
 		1 & \text{if }  |a| > \lambda, \\
 		0 &   -\lambda \le a \le  \lambda . 
 	\end{cases}
 \end{equation}
 where $a \in \mathbb{R}$ is an input signal and $|a|$ is the absolute value of $a$. 
 
\tete{We remark here that it is the convention that logic “1” and “0” are reversed in \eqref{eq:window_detector} and thus the NAND gate in Fig. \ref{fig:WC} becomes AND gate. However, this is not a problem in our case as the role of 1’s and 0’s can be arbitrarily switched  during signal processing with no change on our results. }
 
Consider two zero-mean stationary Gaussian processes 
$y_t \sim \mathcal{N}(0,\sigma_0^2)$ and $\tilde{y}_t \sim \mathcal{N}(0,\sigma_1^2), t = 1,2,\cdots, n$ where $\sigma_1>\sigma_0$. The first process corresponds to the null hypothesis ($\mathcal{H}_0$) and the second to alternative hypothesis ($\mathcal{H}_1$).

\begin{figure*}[!ht]
	\centering
	\includegraphics[width=0.8\textwidth]{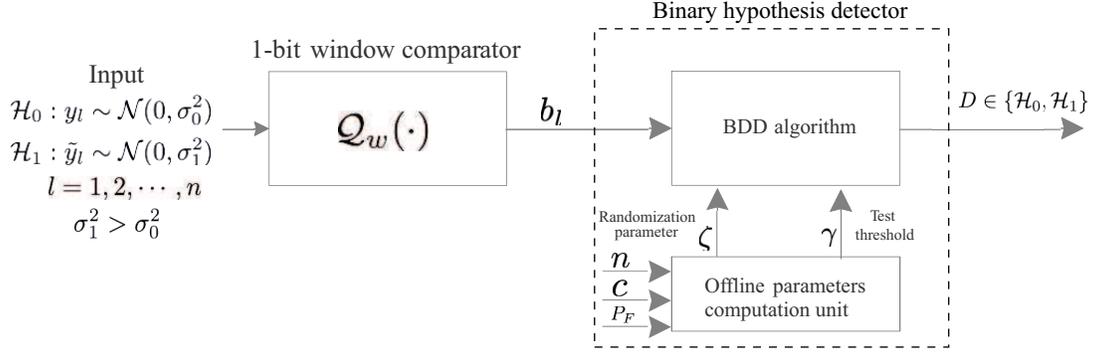}
	\caption{The proposed 1-bit receiver structure for binary hypothesis testing.}
	\label{fig:basic_system_model} 
\end{figure*}

Each process undergoes 1-bit quantization after passing through a 1-bit window comparator defined by \eqref{eq:window_detector}. Thus our binary hypothesis testing problem may be expressed as
 \begin{align}
 	\label{eq:Problem_Formulation}
 	\nonumber
 		\mathcal{H}_0:&  b_1=\mathcal{Q}_w(y_1), \cdots, b_n = \mathcal{Q}_w(y_n),  y_t \sim  \mathcal{N}(0,\sigma_0^2),\\
 		\mathcal{H}_1:&  b_1=\mathcal{Q}_w(\tilde{y}_1), \cdots,b_n= \mathcal{Q}_w(\tilde{y}_n),   \tilde{y}_t\sim  \mathcal{N}(0,\sigma_1^2).
 \end{align}   

The receiver needs to determine if the binary sequence $\{b_1,\cdots,b_n\} \in \{0,1\}^n$ has been caused under $\mathcal{H}_0$ or $\mathcal{H}_1$. \tete{Throughout this work, we assume that $\sigma_0^2$ is known at the receiver, whereas $\sigma_1^2$ is assumed unknown.}

\section{ The proposed 1-bit receiver structure}
\label{sec:section3}

\tete{This section presents the basic 1-bit receiver structure proposed to facilitate binary hypothesis testing. With reference to Fig. \ref{fig:basic_system_model},} the receiver comprises a 1-bit window comparator for quantizing an incoming signal into binary digits, followed by a very low-complexity binary hypothesis detector that runs BDD algorithm. The detector attempts
to decide which hypothesis is most likely to have given rise to the observed binary sequence. We first treat the proposed 1-bit window comparator, then the binary hypothesis detector.

As will be unfolded later, our adapted window comparator makes it possible to recapture a significant amount of information about the presence of an embedded unknown signal in a received waveform.

\begin{figure}[!ht]
	\centering
	\includegraphics[scale=0.5]{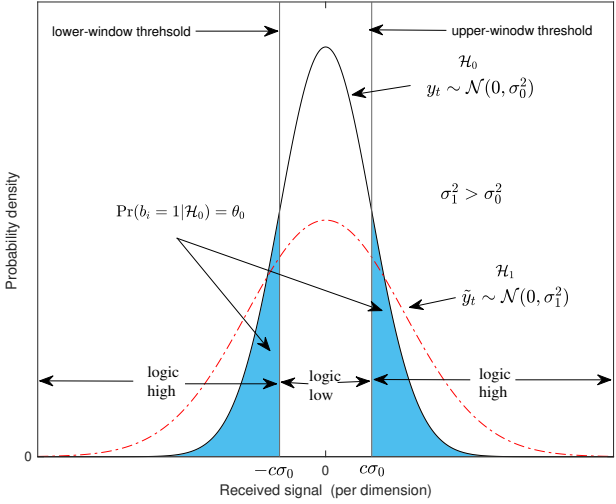}
	\caption{Probabilistic output analysis of window comparator.}
	\label{fig:Fig1} 
\end{figure}
\subsection{The proposed 1-bit window comparator}

We consider the window comparator in Fig. \ref{fig:WC} with $w_U = c \sigma_0$ and $w_L=-c\sigma_0$, where $c$ is a fixed constant (window comparator parameter) which can be optimized to maximize the overall probability of detection. The value of $c$ is typically between 1 and 2, i.e., window width is between two to four times the standard deviation of the model for $\mathcal{H}_0$. In Sec. \ref{sec:section5} we have presented a set of numerical results which suggests that $c \approx 1.6$ serves as an appropriate value, maximizing the detection probability under different requirements of the probability of false alarm. A slight deviation from $c=1.6$ is not very critical, and simply the performance degradation is insignificant. Obviously, letting the window width be too small or too large is useless.

Fig. \ref{fig:Fig1} illustrates the probabilistic output of our window comparator in response to two zero-mean discrete-time Gaussian processes, $y_t \sim \mathcal{N}(0,\sigma_0^2)$ and $\tilde{y}_t\sim \mathcal{N}(0,\sigma_1^2)$. Conditioned on $\mathcal{H}_0$, the window comparator outputs logic 1 with probability given by $\theta_0 = \mathbb{P}(b_i=1|\mathcal{H}_0)$, i.e., null success rate. The shaded area shows this probability in the figure. Conditioned on $\mathcal{H}_1$, the alternative success rate, denoted by $\theta_1=\mathbb{P}(b_i=1|\mathcal{H}_1)$ is given by the area under the dashed curve in the intervals $(c \sigma_0, \infty)$ and $(-\infty, -c \sigma_0)$, where this area increases monotonically with increasing $\sigma_1>\sigma_0$. 

Referring to Fig. \ref{fig:Fig1}, we have that  
\begin{eqnarray}
	\label{eq:theta0}
	\nonumber
	\theta_0 &= &  2 \int_{c\sigma_0}^{\infty} \frac{1}{\sqrt{2 \pi \sigma_0^2}} \exp \left[{-\frac{y^2}{2 \sigma_0^2}}\right]dy \\
	&=&	2 Q(c)
\end{eqnarray} 
\begin{eqnarray}
	\label{eq:theta1}
	\nonumber
	\theta_1 & =& 2 \int_{c\sigma_0}^{\infty} \frac{1}{\sqrt{2 \pi \sigma_1^2}} \exp\left[{-\frac{y^2}{2 \sigma_1^2}}\right]dy \\
	&=&	2 Q\left(c\frac{\sigma_0}{\sigma_1}\right) \triangleq  2 Q(\alpha c  )
\end{eqnarray} 
\titi{where $Q(\cdot)$ is the complementary distribution function (aka. Q-function) of a standard normal random variable} and $\alpha$ is defined by 
\begin{equation}
	\label{eq:alpha}
	\alpha = {\sigma_0}/{\sigma_1}   \in (0,1).
\end{equation}

\tete{From \eqref{eq:theta0} and \eqref{eq:theta1}, it is worth remarking that  $\theta_0$ is always fixed regardless of $\sigma_0^2$ and $\theta_1$ being monotonically increasing with increasing ratio $\alpha \in (0,1)$ is the key feature of our quantization method for an efficient detection process. }

Again, in the light of \eqref{eq:theta0} and \eqref{eq:theta1}, an observed binary sequence $b_1,\cdots, b_n$ at the output of our window comparator is viewed as a realization of an i.i.d. Bernoulli process with success rate $\theta_0$ (known and fixed) under $\mathcal{H}_0$ or with success rate $\theta_1$ (unknown) under $\mathcal{H}_1$. Asymptotically, the number of 1’s in the binary sequences generated under $\mathcal{H}_0$ and $\mathcal{H}_1$ will roughly be $2nQ(c)$ and $2nQ(\alpha c) >2nQ(c)$, respectively. \tete{This means that, in the asymptotic sense, we have two distinct sets of binary sequences, but in practice, $n$ is usually finite, therefore there will be some overlapping between these two sets giving rise to detection confusion.}

\subsection{Test statistic}
The problem now has been reduced to a simple one; to distinguish between two binary sequences of length $n$, generated by two Bernoulli processes with $n$ trials and different success rates. That is,
\begin{align} 
	\label{eq:Bernoulli_Hypothesis}
	\nonumber
	\mathcal{H}_0 &: 	b_1,\cdots,b_n\overset{i.i.d.}{\sim} \operatorname{Ber}(\theta_0)\\
	\mathcal{H}_1 &: 		b_1,\cdots,b_n\overset{i.i.d.}{\sim} \operatorname{Ber}(\theta_1)
\end{align}
where the notation $b \sim \operatorname{Ber}(\theta)$ is a shorthand for the random variable $b$ being distributed according to a Bernoulli probability mass function(PMF) of success rate $\theta$, i.e., $\mathbb{P}(b)= \theta^{b} (1-\theta)^{1-b}, b\in \{0,1\}$. Thus the joint PMFs under both hypotheses are \cite{gubner_2006},  
\begin{align} 
	\label{eq:H0_quantized}
	\nonumber
	\mathcal{H}_0 &: 	\mathbb{P}(b_1,b_2,\cdots,b_n|\theta_0) = \prod_{i=1}^{n} \theta_0^{b_i} (1-\theta_0)^{1-b_i}\\
	\mathcal{H}_1 &: 	\mathbb{P}(b_1,b_2,\cdots,b_n|\theta_1) = \prod_{i=1}^{n} \theta_1^{b_i} (1-\theta_1)^{1-b_i},
\end{align}
 where $\mathbb{P}(\cdot|\theta)$ denote joint PMF parameterized by $\theta$. 
 
 Using the log-likelihood ratio (LLR) \cite{detection95} as a sufficient statistic yields
 
\begin{align}
	\label{eq:LLR_quantized0}
	\nonumber
	& {\Lambda} (\{b_i\}) =  \log \left [ \frac{ \mathbb{P}(b_1,b_2,\cdots,b_n|\theta_1)}{\mathbb{P}(b_1,b_2,\cdots,b_n|\theta_0)} \right] \\\nonumber
	&= \log \left[  \frac{ \prod_{i=1}^{n} \theta_1^{b_i} (1-\theta_1)^{1-b_i}}{\prod_{i=1}^{n} \theta_0^{b_i} (1-\theta_0)^{1-b_i}} \right]  \\
&= \sum_{i=1}^{n} b_i    {\log \left[  \frac{\theta_1 (1-\theta_0)}{\theta_0 (1-\theta_1)} \right]}+ n \log \left [  \frac{1-\theta_1}{1-\theta_0} \right] \titi{\quad\mathop{\gtrless}_{D  =\mathcal{H}_0}^{D=\mathcal{H}_1}} \gamma^{\prime}
\end{align}
where $\gamma^{\prime}$ is an arbitrary test threshold. \tete{The notation $\quad\mathop{\gtrless}_{D  =\mathcal{H}_0}^{D=\mathcal{H}_1}$ in the last line of \eqref{eq:LLR_quantized0} means that the LLR decision is that $D = \mathcal{H}_1$ if the left-hand side is greater than $\gamma^{\prime}$, and is that $D= \mathcal{H}_0$ if the left-hand side is less than $\gamma^{\prime}$.}

From the third line of \eqref{eq:LLR_quantized0}, taking the second term in the left-hand side to the right-hand side, we obtain
\begin{equation}
	\label{eq:LLR_quantized00}
	\sum_{i=1}^{n} b_i    \log \left[  \frac{\theta_1 (1-\theta_0)}{\theta_0 (1-\theta_1)} \right]  \quad\mathop{\gtrless}_{D=\mathcal{H}_0}^{D=\mathcal{H}_1} \gamma^{\prime}-n \log \left[  \frac{1-\theta_1}{1-\theta_0} \right]
\end{equation}

Since $\theta_1> \theta_0$, the right-hand side of \eqref{eq:LLR_quantized00} and the multiplicative factor, $\log \left[  {\theta_1 (1-\theta_0)}/{\theta_0 (1-\theta_1)} \right]$, on the left-hand side are nonnegative numbers. Also, the factor, $\log \left[  {\theta_1 (1-\theta_0)}/{\theta_0 (1-\theta_1)} \right]$, is not a function of $b_i$'s. Therefore, dividing both sides of \eqref{eq:LLR_quantized00} by $\log \left[  {\theta_1 (1-\theta_0)}/{\theta_0 (1-\theta_1)} \right]$,  \eqref{eq:LLR_quantized00} boils down to the following sufficient statistic test:
\begin{equation}
	\label{eq:LLR_quantized}
	\mathcal{S}_n=  \sum_{i=1}^{n} b_i \quad\mathop{\gtrless}_{D = \mathcal{H}_0}^{D=\mathcal{H}_1}\gamma
\end{equation}
where $\gamma$ is the modified detection threshold given by
\begin{equation}
	\gamma \triangleq   \left (  {\gamma^{\prime} -n \log \left[ \frac{1-\theta_1}{1-\theta_0}\right]} \right) \bigg / {\log \left[  \frac{\theta_1 (1-\theta_0)}{\theta_0 (1-\theta_1)}   \right]}.
\end{equation}

\tete{Note that \eqref{eq:LLR_quantized} partitions the space of observed sequences based on the number of 1's into two regions.The first region is $\mathcal{R}_1: \{ \{b_i\}: \mathcal{S}_n=\sum_{i=1}^{n}b_i> \gamma\}$ where the receiver accepts $\mathcal{H}_1$ as true and rejects $\mathcal{H}_0$, and the second region is $\mathcal{R}_2: \{ \{b_i\}: \mathcal{S}_n=\sum_{i=1}^{n}b_i < \gamma\}$  where the receiver accepts $\mathcal{H}_0$ as true and rejects $\mathcal{H}_1$. }

\tete{We see that given $\{b_i\}$, the sufficient statistic in \eqref{eq:LLR_quantized} does not depend on the unknown parameter $\theta_1$ ($\theta_1, \theta_0$ and $\gamma^{\prime}$ are all absorbed in $\gamma$ which is a constant to be chosen at our disposal)} and is monotonically increasing in the sense that, the more 1’s in the observed binary sequence, the more likely $\mathcal{H}_1$ will be. According to the \emph{Neyman-Pearson criterion}, it is a uniformly most powerful (UMP) \cite{detection95} test in the sense that for any value of $\theta_1$ (or equivalently $\sigma_1$), the resulting probability of detection, $P_D$, is maximized for a fixed probability of false alarm $P_F$. The threshold $\gamma$ is determined by the distribution of $\mathcal{S}_n$.

\subsection{The proposed BDD algorithm}
\label{subsec:Analysis of BDD algorithm}
Since the null success rate, $\theta_0 = 2Q(c)$\eqref{eq:theta0}, is available at the receiver beforehand, the test threshold $\gamma$ can be selected in away such that the resulting probability of false alarm is less than or equal to a prescribed probability of false alarm, $P_F$. From \eqref{eq:LLR_quantized}, the probability of false alarm results from accepting $\mathcal{H}_1$ when $\mathcal{H}_0$ was true, i.e., $\mathbb{P}(D=\mathcal{H}_1|\mathcal{H}_0) = \mathbb{P}(\mathcal{S}_n>\gamma|\mathcal{H}_0)$. Thus it is determined by the distribution of the sufficient statistic, $\mathcal{S}_n$.

To proceed, we note that $\mathcal{S}_n$ is a sum of i.i.d. Bernoulli random variables, hence it follows the \emph{binomial} distribution under both hypotheses:
\begin{align} 
	\label{eq:BinomialPDF}
	\nonumber
	\mathcal{H}_0&: 	     \mathcal{S}_n \sim \operatorname{Bin}(n, \theta_0) \\ \mathcal{H}_1 &: 	  \mathcal{S}_n \sim  \operatorname{Bin}(n, \theta_1).
\end{align}

\titi{Therefore, $\gamma$ can be selected such that }
\begin{align}
	\label{eq:PA_true}
	\nonumber
	 P_F& \titi{\ge}  \mathbb{P}( \mathcal{S}_n >\gamma|\mathcal{H}_0) \\ \nonumber
	&= \sum_{l=\gamma+1}^{n}\binom{n}{l}  (2 Q(c))^{l} (1-2Q(c))^{n-l} \\
	&=1-\sum_{l=0}^{\gamma}\binom{n}{l}  (2 Q(c))^{l} (1-2Q(c))^{n-l}
\end{align}
 \tete{ and rearranging \eqref{eq:PA_true} yields
 \begin{equation}
 	\label{eq:PA_true2}
 	\underbrace{ \sum_{l=0}^{\gamma}\binom{n}{l}  (2 Q(c))^{l} (1-2Q(c))^{n-l}}_{\mbox{Binomial CDF}} \ge \underbrace{ 1-P_F}_ {\triangleq \bar{P}_F}.
 \end{equation}}
 
 The summation on the left side of \eqref{eq:PA_true2} can be easily computed numerically to find $\gamma$. \tete{For notational  convenience,  $\gamma$ can be  expressed} by the inverse binomial cumulative distribution function (CDF), i.e.,
\begin{equation}
	\label{eq:InverseCDF} 
	\gamma = \operatorname{Bin}^{-1}(\bar{P}_F,n, 2Q(c))
\end{equation}
\tete{where, in the light of \eqref{eq:PA_true2}, the inverse binomial CDF, $\operatorname{Bin}^{-1}(\cdot)$, returns the smallest integer $\gamma$ such that the binomial CDF evaluated at $\gamma$ is equal to or exceeds $\bar{P}_F$.}

A closed-form solution for \eqref{eq:InverseCDF} is indeed intractable. Nevertheless, an approximate solution for $\gamma$ is possible by invoking the central limit theorem when $n$ is sufficiently large. This allows us to approximate the binomial distribution by a Gaussian distribution. That is, under $\mathcal{H}_0$, the sum $\mathcal{S}_n$ of observed $n$ bits can be approximated as a Gaussian random variable with mean $n\theta_0$ and variance $n\theta_0 (1-\theta_0)$, i.e., 
$$\mathcal{S}_n \sim \mathcal{N}(n\theta_0, n\theta_0 (1-\theta_0)).$$ Then approximately,
\begin{align}
	\label{eq:P_F_GaussianApprox}
	\nonumber
	P_F &\approx  \int_{\gamma}^{\infty} \frac{1}{\sqrt{2 \pi n\theta_0 (1-\theta_0)}} \exp\left[-\frac{(u-n\theta_0)^2}{2 n\theta_0 (1-\theta_0)}\right]du \\
	&=Q \left (\frac{\gamma-n\theta_0}{\sqrt{n\theta_0 (1-\theta_0)}} \right) = Q \left (\frac{\gamma-2 n Q(c)}{\sqrt{2 n Q(c) (1-2Q(c))}} \right) 
\end{align}
and hence $\gamma$ can be obtained as the smallest integer given by 
\begin{equation}
	\gamma \approx \floor {\sqrt{2nQ(c) (1-2Q(c))}Q^{-1}(P_F) + 2 n Q(c)}
\end{equation}
where $Q^{-1}(\cdot)$ denotes the inverse Q-function of standard normal distribution and $\floor {x}$ denotes the floor function which returns the smallest integer which is less than or equal to $x$.

In the above analysis, note that in addition to $n$ being sufficiently large, the Gaussian approximation is acceptable when $n\theta_0(1-\theta_0)$ is not too small, which is not a problem in our case.



Unlike a continuous distribution, binomial distribution is discrete. Hence, it is usually the case that there is no integer-valued $\gamma$ that satisfies \eqref{eq:PA_true} with equality and hence the closest solution is accepted, resulting in a probability of a false alarm that is less than the desired one. Also, note that there is a nonzero probability for the event $\mathcal{S}_n = \gamma$ which is not included in the test, in contrast to a continuous random variable where the probability at a specific value is zero. Therefore, the resulting probability of false alarm due to accepting the closest solution mentioned above will change with $n$, \titi{which affects the desired probability of detection negatively and hence we propose to avoid this situation as explained next}.  

\titi{Thus, we propose }to maintain the probability of false alarm at the desired level $P_F$ by introducing some \titi{controlled} randomization in the decision process, according to the following rule:
	\begin{equation} 
		\label{eq:randomization}
		\operatorname{Decide:}
	\begin{cases}
	\mathcal{H}_1 &\text{if  } \mathcal{S}_n > \gamma  \\
	\mathcal{H}_0 &\text{if  } \mathcal{S}_n < \gamma  \\
	\mathcal{H}_1 \text{ with prob. } \zeta & \text{if  } \mathcal{S}_n = \gamma  \\
		\mathcal{H}_0 \text{ with prob. } 1-\zeta & \text{if  } \mathcal{S}_n = \gamma.
	\end{cases}
\end{equation}  
where $\zeta \in (0,1)$ is referred to as the randomization parameter.

According to \eqref{eq:randomization}, the detection algorithm resorts to a randomized decision only when the event $\mathcal{S}_n=\gamma$ occurs. This event occurs with probability $\binom{n}{\gamma}  (2Q(c))^{\gamma} (1-2Q(c))^{n-\gamma}$ which is very small, compared with the probability of the non-randomized decision. It is worth noting that introducing some sort of a controlled randomization in a decision process is not new and has been proven useful in a variety of well-known algorithms including Markov Chain Monte Carlo (MCMC) based MIMO detector \cite{MCMC} and stochastic population-based search algorithms \cite{Spall2003}.

Based on \eqref{eq:randomization}, \eqref{eq:PA_true} is modified as 
\begin{equation}
	\label{eq:Probability_increment}
P_F = {\mathbb{P}( \mathcal{S}_n >\gamma|\mathcal{H}_0) }+ {\zeta \mathbb{P}( \mathcal{S}_n =\gamma|\mathcal{H}_0)  }
\end{equation}
where the first term on the right side of \eqref{eq:Probability_increment} corresponds to the probability of false alarm due to non-randomized decision (i.e., LLR decision \eqref{eq:LLR_quantized00}) and the second term to the randomized decision (i.e., when the event $\mathcal{S}_n =\gamma$ occurs).  From \eqref{eq:Probability_increment}, we have that
\begin{align}
	\label{eq:zeta}
	\nonumber
	\zeta &= \frac{ P_F-\mathbb{P}( \mathcal{S}_n >\gamma|\mathcal{H}_0)}{\mathbb{P}( \mathcal{S}_n =\gamma|\mathcal{H}_0)} \\
	&= \frac{P_F - \sum_{l=\gamma+1}^{n} \binom{n}{l}  \left(2 Q(c) \right)^{l}  \left(1-2 Q(c  ) \right)^{n-l}}{ \binom{n}{\gamma}  \left(2 Q(c) \right)^{\gamma}  \left(1-2 Q(c  ) \right)^{n-\gamma}} 
\end{align}
which is computed offline at the receiver as all involved parameters in \eqref{eq:zeta} are known beforehand.

With the aid of \eqref{eq:Probability_increment}, it is easy to note that when $P_F \to 0 (\gamma=n)$, $\zeta \to 0$ and when $P_F \to 1 (\gamma=0)$, $\zeta \to 1$ and hence $\zeta \in [0,1]$, verifying that $\zeta$ is a proper probability value. The BDD algorithm is summarized in Algorithm \ref{alg:alg1}.
\begin{algorithm}
	\caption{\centering{BDD algorithm}}
	\begin{algorithmic}
		\State \textbf{Initialization:} $n,c, P_F$
			\State \textbf{Input:} $\gamma$ \eqref{eq:InverseCDF}, $\zeta$ \eqref{eq:zeta}$\{ b_1,b_2,\cdots,b_n\} \in \{0,1\}^n$
		\State \textbf{Test statistic:} $\mathcal{S}_n = \sum_{i=1}^{n} b_i$
	\State \textbf{Result:}  $D \in \{\mathcal{H}_0,\mathcal{H}_1\}$
	\If{ $S_n=\gamma$} 
	\State {Generate:  $X \sim \operatorname{Uniform} (0,1)$}
$$D = 
	\begin{cases}
			\mathcal{H}_1 &\text{ if  }  X  < \zeta  \\
			\mathcal{H}_0 &\text{ if  } X> \zeta  \\
				\end{cases}$$
\Else
$$D = 
	\begin{cases}
		\mathcal{H}_1 &\text{ if  }  \mathcal{S}_n  > \gamma  \\
		\mathcal{H}_0 &\text{ if  } \mathcal{S}_n<\gamma  \\
	\end{cases}$$
\EndIf
	\end{algorithmic}
	\label{alg:alg1}
\end{algorithm}

\tete{\textbf{\emph{Summary of Algorithm \ref{alg:alg1}}:} The initial parameters are the number of binary observations, $n$, window comparator parameter, $c$, the desired probability of false alarm, $P_F$. The main inputs used by the algorithm are the test threshold $\gamma$, a randomization parameter $\zeta$ and the observed sequence (see also Fig. \ref{fig:basic_system_model}). $\gamma$ and $\zeta$ are computed offline in order using \eqref{eq:InverseCDF} and \eqref{eq:zeta}, respectively. Note that $n,c,P_F,\gamma,\zeta$ are fixed and known beforehand. Then, the algorithm computes the number of 1’s in the received sequence (i.e., $\mathcal{S}_n=\sum_{i=1}^{n} b_i$) and decides $\mathcal{H}_0$ or $\mathcal{H}_1$ using one of two approaches: randomized test and LLR test (i.e., binomial test). The decision is based on the first approach if $\mathcal{S}_n = \gamma$, but on the second approach if $\mathcal{S}_n \neq \gamma$. In the randomized test, $\mathcal{H}_1$ is decided with probability $\zeta$ whereas $\mathcal{H}_0$ with probability $1-\zeta$. To achieve these probabilities, we generate a uniform random number $X \in (0,1)$ and compare it with $\zeta$. If $X<\zeta$, then $\mathcal{H}_1$ is selected, or $\mathcal{H}_0$ is selected otherwise.  }

The probability of detection $P_D$ is given in the following result.
\begin{thm}
		\label{thm:PD_WD}
For a fixed probability of false alarm $P_F$, the probability of detection of BDD algorithm is
\begin{align}
		\label{eq:P_D_true_BDJD}
	\nonumber
P_D &= \sum_{l=\gamma+1}^{n} \binom{n}{l}  \left(2 Q(\alpha c) \right)^{l}  \left(1-2 Q(\alpha c  ) \right)^{n-l}  \\ 
&+  \frac{P_F - \sum_{l=\gamma+1}^{n} \binom{n}{l}  \left(2 Q(c) \right)^{l}  \left(1-2 Q(c  ) \right)^{n-l}}{ \left(\frac{Q(c)}{Q(\alpha c)} \right)^{\gamma}  \left(  \frac{1-2 Q(c )}{1-2Q(\alpha c)} \right)^{n-\gamma}}. 
\end{align}
\end{thm}
\begin{proof}
The result follows directly from \eqref{eq:randomization} by noting that $$P_D =  \mathbb{P} ( \mathcal{S}_n > \gamma|\mathcal{H}_1) +  \zeta \mathbb{P} ( \mathcal{S}_n =\gamma|\mathcal{H}_1)$$ where $\zeta$ is given in \eqref{eq:zeta}, $\mathbb{P} ( \mathcal{S}_n =\gamma|\mathcal{H}_1)$ and $\mathbb{P} ( \mathcal{S}_n > \gamma|\mathcal{H}_1)$ are respectively defined by
\begin{align*}
\mathbb{P} ( \mathcal{S}_n =\gamma|\mathcal{H}_1)&=  \binom{n}{\gamma}  \left(2 Q(\alpha c) \right)^{\gamma}  \left(1-2 Q(\alpha c ) \right)^{n-\gamma} \\
\mathbb{P} ( \mathcal{S}_n > \gamma|\mathcal{H}_1) &= \sum_{l=\gamma+1}^{n} \binom{n}{l}  \left(2 Q(\alpha c) \right)^{l}  \left(1-2 Q(\alpha c  ) \right)^{n-l}.
	\end{align*}
\end{proof}

\tete{In \eqref{eq:P_D_true_BDJD}, the first and second terms on the right side correspond to the detection probability resulting from the non-randomized (LLR) and randomized decisions, respectively.}

The following result shows the asymptotic behaviour of $P_D$ when the number of observations grows large or/and the change in variance between the two hypotheses becomes sufficiently large.
\begin{cor}
	\label{cor:Asymptotic_in_alpha_and_n}
	When $\alpha\to 0$ or/and $n\to \infty$, $P_D$ converges to 1,
	\begin{equation}
		\lim\limits_{\alpha \to 0} P_D = \lim\limits_{n\to \infty} P_D = 1
	\end{equation}
\end{cor}
\begin{proof}
The result follows directly by substituting $\alpha=0$ or $n=\infty$ in \eqref{eq:P_D_true_BDJD} while other parameters are held fixed.
\end{proof}
In practice, Corollary \ref{cor:Asymptotic_in_alpha_and_n} is valid  for modest values of $\alpha$ and $n$. 

The probability of detection, $P_D$, for a traditional zero-threshold 1-bit quantizer is recovered by substituting $c=0$ (the upper- and lower-window thresholds collapse into a single zero-threshold) in \eqref{eq:P_D_true_BDJD}. The resulting $P_D$ is given in the following result.
\begin{cor}
		\label{cor:PD_conventional_quantizer}
	Under a traditional zero-threshold 1-bit quantizer, $P_D$ converges to $P_F$, i.e.,
	\begin{equation}
		\lim_{c\to 0} P_D = P_F
	\end{equation}
\end{cor}

This result shows that, under the conditions considered in this work, the traditional 1-bit quantizer fails completely to distinguish between the two hypotheses. We give next a remark on the optimal value of $c$.

\begin{remark}
	\label{rem:remark1}
Note that obtaining the optimal value of $c$ that maximizes \eqref{eq:P_D_true_BDJD} does not lead to a closed form solution. \titi{In general, the value of $c$ can be chosen between 1 and 2, i.e., one to two times the standard deviation of the model in $\mathcal{H}_0$}. The optimal $c$ can be simply computed numerically \titi{which is found to be approximately $c\approx 1.6$ when change in variance between the models in $\mathcal{H}_0$ and $\mathcal{H}_1$ is very small}, as discussed in Sec. \ref{sec:section5}.

Note that for the two extremes ($c=0$ and $c=\infty$)---which are not of interest---the probability of detection \eqref{eq:P_D_true_BDJD} converges to $P_F$, 
	$$\lim_{c\to \infty} P_D =\lim_{c\to 0} P_D \to P_F.$$
\end{remark}

\subsection{Transmission of binary observations via a noisy channel}
If the binary observations $b_1,\cdots,b_n$ need to be transmitted over a noisy channel to another party responsible for taking the final decision, then the test statistic needs to be changed to account for the error introduced by the channel. 

Assume a binary symmetric channel (BSC) with crossover probability $\epsilon$. Then \eqref{eq:BinomialPDF} becomes 
\begin{align} 
	\label{eq:BinomialPDF_BSC}
	\nonumber
	\mathcal{H}_0&: 	  \mathcal{S}_n \sim \operatorname{Bin}(n,  \theta_0^{\prime})\\
	\mathcal{H}_1&: 	  \mathcal{S}_n \sim  \operatorname{Bin}(n,\theta_1^{\prime}).
\end{align}
where $\theta_0^{\prime}=\epsilon(1-2Q(c))+2(1-\epsilon) Q(c)$ and $\theta_1^{\prime}=\epsilon(1-2Q(\alpha c))+2(1-\epsilon) Q(\alpha c)$ are the new success rates of the Bernoulli random variables after transmission via the BSC.

Thus, replacing $\theta_0 = 2Q(c)$ and $\theta_1 = 2Q( \alpha c)$ by $\theta_0^{\prime}$ and $\theta_1^{\prime}$ in \eqref{eq:InverseCDF}, a new test threshold $\gamma^{\text{BSC}}$ is obtained. The new detection probability in Theorem \ref{thm:PD_WD} is thus modified as
\begin{align}
	\label{eq:P_D_true_BDJD_BSC}
	\nonumber
	P_D^{\text{BSC}} &= \sum_{l=\gamma^{\text{BSC}}+1}^{n} \binom{n}{l}  \left(\theta_1^{\prime}\right)^{l}  \left(1-\theta_1^{\prime} \right)^{n-l}  \\ 
	&+  \frac{P_F - \sum_{l=\gamma^{\text{BSC}}+1}^{n} \binom{n}{l}  \left(\theta_0^{\prime}  \right)^{l}  \left(1-\theta_0^{\prime} \right)^{n-l}}{ \left(\frac{\theta_0^{\prime} }{\theta_1^{\prime}} \right)^{\gamma^{\prime}}  \left(  \frac{1-\theta_0^{\prime}}{1-\theta_1^{\prime}} \right)^{n-\gamma^{\prime}}}.
\end{align}

In practice a good channel coding is used which can reduce the error introduced by the channel significantly, therefore, the (error-free) detection probability in Theorem \ref{thm:PD_WD} is still achievable.

\section{Applications}
\label{sec:section4}

In this section, \tete{we consider the straightforward extension of the proposed 1-bit receiver structure in Fig. \ref{fig:basic_system_model}} for two MIMO wireless applications. The first is jamming detection in a massive MIMO system, and the second is signal probing in a WSN. In both cases, we assume that all channels follow \emph{Rayleigh fading} with unknown instantaneous realizations at receivers.

\subsection{Massive MIMO}  
\begin{figure*}[!ht]
	    \centering
	\includegraphics[width=0.75\textwidth]{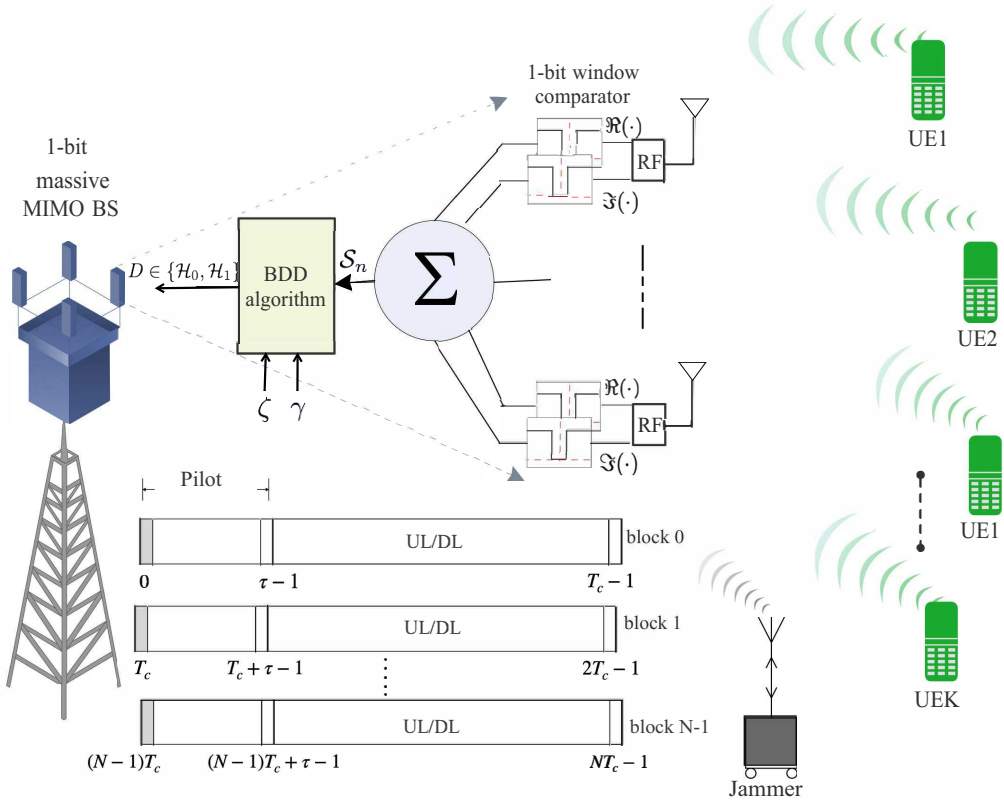}
	\caption{An uplink massive MIMO system with BS equipped with a set of 1-bit window comparators and BDD algorithm for detecting a pilot attack. In this example, the first pilot symbol of each coherence channel block is used over $N$ independent channel blocks. }
	\label{fig:BDD_massiveMIMO} 
\end{figure*}

One-bit quantized massive MIMO system, where the BS has only access to binary measurements, is desirable for its low-hardware complexity and insignificant energy consumption. Fig. \ref{fig:BDD_massiveMIMO} shows a conceptual schematic diagram of a massive MIMO system under pilot attack. \tete{The issue of practical importance here is simply that of detecting a malicious attack by the BS based on coarsely quantized samples (binary samples).}

We propose to equip the BS with a sufficient number of our 1-bit window comparators (not necessarily all antennas) and let the BS run BDD algorithm (Algorithm \ref{alg:alg1}) to detect an unknown jamming in the system, which proves to be a very promising solution for single and multiple-antenna jamming devices.

 \titi{In our model, all channels are assumed \emph{flat Rayleigh-block fading} where all channel realizations are unknown at the BS, the large-scale fading coefficients and transmit power of UEs are also known at the BS, however, nothing is known about the jammer, except its channel statistics}. Note that the assumption of channel being unknown is more realistic than the assumption of perfect channel, especially when a signal is coarsely quantized  and (unknown) jamming power is high, which may render the channel estimate useless.

\paragraph{\textbf{Signal model and problem formulation}}
We consider the uplink (training phase) in a single-cell  multiuser MIMO system operating in time-division duplex (TDD) mode, i.e., see  Fig. \ref{fig:BDD_massiveMIMO}. The system has $K$ users served by a BS with $M$ antennas  in the same time-frequency resource. Without loss of generality, we consider a single-antenna jammer who sends the same pilot of user $k$ (i.e., pilot spoofing) to contaminate its channel estimate at the BS. The case of a multiple-antenna jammer transmitting arbitrary pilot signal is also workable \tete{as will be shown in Sec. \ref{sec:section5}. }

We use $\vv{h}_k, \vv{g} \sim \mathcal{CN} (\vv{0},\vv{I}_M)$  to denote channel vectors  (Rayleigh small-scale fading) corresponding to user $k$ and the jammer, respectively, and all channel vectors stay constant over $T_c$ symbols (i.e., coherence time). 

With no loss of generality and for notational simplicity, we equip all $M$ antennas with 1-bit window comparators, one per signal dimension. Thus, under $\mathcal{H}_0$ (i.e., no jamming) and $\mathcal{H}_1$(i.e., with jamming), the unquantized baseband discrete-time signal received at the BS at time $t$ is
\begin{align}
	\label{eq:unquantized_rx}
	\nonumber
\mathcal{H}_0:	\vv{y}_t &=  \sum_{i=1}^{K} \sqrt{ \beta_i p_i } \vv{h}_i {s}_{i,t} + \vv{w}_t \\
\mathcal{H}_1: \vv{y}_t&= \sum_{i=1}^{K} \sqrt{ \beta_i p_i } \vv{h}_i {s}_{i,t}+ \sqrt{\beta_J p_J } \vv{g}  {s}_{k,t} + \vv{w}_t 
\end{align}
where $s_{i,t}$ is the pilot symbol sent from the $i$-th user at time instant $t$, $p_i$ and $p_J$ denote the average transmit powers of user $i$ and jammer, $\beta_i$ and $\beta_J$ are the large-scale coefficients (e.g., path loss and shading) of user $i$ and jammer, respectively, and $\vv{w}_t \sim \mathcal{CN} (\vv{0},\vv{I}_M)$ is uncorrelated white Gaussian noise. Without loss of generality, we assume $|s_{i,t}|=1$, i.e., constant modulus pilot symbols.

In the following analysis, we study a case where jamming detection is performed using only one pilot symbol (e.g., first symbol) received per coherence time and $N$ coherence blocks (independent realizations of channels) is considered as illustrated in Fig. \ref{fig:BDD_massiveMIMO}. It is worth noting that jamming detection using all pilot symbols during one coherence time is possible as long as $K=\tau$ for the received samples being independent (with channel being random) under $\mathcal{H}_0$ and hence our test is computable. Otherwise correlation should be taken into account leading to uncomputable test statistic, due to the intractability of distribution of correlated Bernoulli random variables.

\paragraph{\textbf{Jamming detection}}
According to our channel model, conditioned on each hypothesis, the unquantized signal $\vv{y}_t$ is complex Gaussian with i.i.d. entries, i.e., 
\begin{align}
	\label{eq:unquantized_rx_covmatrix}
	\nonumber
	\mathcal{H}_0 &:	\vv{y}_t \sim \mathcal{CN} \left(\vv{0}, \sigma_0^2 \vv{I}_M \right)\\ 
	 \mathcal{H}_1 &:	\vv{y}_t \sim \mathcal{CN} \left(\vv{0}, \sigma_1^2 \vv{I}_M \right)\\ \nonumber
	t&=0,T_c,2T_c,\cdots, (N-1)T_c
\end{align}
where   $N \ge 1$ is the number of channel blocks and $\sigma_0^2, \sigma_1^2$ are given by
\begin{align}
	\label{eq:var_massivemimo}
\sigma_0^2 = \sum_{i=1}^{K} p_i \beta_i  +\sigma_w^2, \hspace{1em}
\sigma_1^2 =\sum_{i=1}^{K} p_i \beta_i  + p_J \beta_J + \sigma_w^2.
\end{align}
\tete{We recall that $\sigma_0^2$ is assumed known at the BS, whereas $\sigma_1^2$ is unknown. }

Define $\vv{Y} = [\vv{y}_0, \vv{y}_{T_c}, \cdots, \vv{y}_{(N-1)T_c}] \in \mathbb{C}^{M \times N}$ and
$\bar{\vv{y}} = \operatorname{Vec} \left( \left [ \Re\{\vv{Y} \}  , \Im\{\vv{Y} \} \right] \right) \in \mathbb{R}^{2MN}
$ where $\operatorname{Vec}(\cdot)$ denotes the vectorization operator, i.e., we stack all real and imaginary components of $\vv{Y}$ in a column vector. All entries of $\bar{\vv{y}}$ are i.i.d.  $\mathcal{N} (0,\sigma_0^2/2)$ under $\mathcal{H}_0$ and i.i.d. $\mathcal{N} (0,\sigma_1^2/2)$ under $\mathcal{H}_1$.

Let $\vv{b} \triangleq  \mathcal{Q}_w(\bar{\vv{y}}) = [b_1, b_2,\cdots,b_{2 M N }]^{\tran}$ be length-$2MN$ binary sequence oberved after all window comparators at the BS where $\mathcal{Q}_w(\cdot)$ is a component-wise operator. According to our well-designed window comparator, the binary sequence $b_1, b_2,\cdots,b_{2MN}$ is a realization of an i.i.d. Bernoulli random process with different success rates under both hypotheses. Therefore, the jamming detection can be formulated by the following binary hypotheses:
\begin{align} 
	\label{eq:BinaryHypothesesMassiveMIMO}
	\nonumber
	\mathcal{H}_0 &: 	b_1,\cdots,b_{2 M N} \overset{i.i.d.} {\sim} \operatorname{Ber}(2Q(c)) \\
	\mathcal{H}_1 &: b_1,\cdots,b_{2 M N} \overset{i.i.d.} {\sim}  \operatorname{Ber}(2Q(c \sigma_0 /\sigma_1)) 
\end{align} 
where $\sigma_0$ and $\sigma_1$ are defined in \eqref{eq:var_massivemimo}.

Given \eqref{eq:BinaryHypothesesMassiveMIMO}, the BS then runs BDD algorithm (Algorithm \ref{alg:alg1}) to decide if there is jamming or not.

\subsection{Wireless sensor network}
\label{sub:WSN}
WSNs are characterized by limited resources such as bandwidth and energy. Because of such system constraints, observations at cheap sensors in the network require to be compressed before being sent to a central unit for processing. Particularly, when the constraint on energy and bandwidth is severe, the sensors map the observations into binary measurements before transmission. 

\begin{figure*}[!htp]
	\centering
	\includegraphics[width=0.5\textwidth]{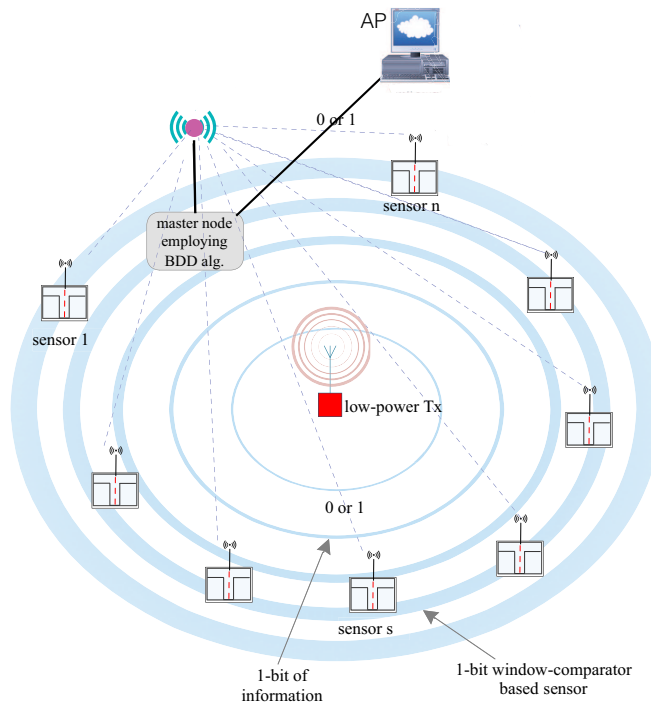}
	\caption{ A wireless sensor network employing 1-bit window comparator at all sensors except the master sensor, to detect the presence or absence of a low-power signal against background noise. The transmitter sends only one symbol ($\tau = 1$) and each sensor reports its observed bit to the master node for the final decision.}
	\label{fig:WSN_schematic} 
\end{figure*}

A conceptual schematic diagram of a WSN scenario is shown in Fig. \ref{fig:WSN_schematic}, where we are interested in probing a signal sent from a low-power transmitter in the neighborhood by a set of cheap sensors employing the similar 1-bit window comparators to quantize the received signals before sending to a master sensor or node. Then the master node takes a decision using our proposed BDD algorithm and sends the result to a central unit (e.g., access point).

\paragraph{\textbf{Signal model and problem formulation}} 
With reference to Fig. \ref{fig:WSN_schematic}, we consider a WSN comprising $n$ spatially distributed sensors where each is equipped with a simple window comparator. We also assume that there is a more capable $\emph{master sensor}$ which collects data from all sensors, running BDD algorithm and reporting the result to an \emph{access point} (AP).

There is no loss of generality in assuming a real-valued signal model where the complex-valued signal can be decomposed into two i.i.d. real-valued random variables under $\mathcal{H}_0$. Each sensor receives (per channel use) a noisy version of a  constant signal, say, ${x} = \sqrt{p}$ sent from a non-stationary low-power transmitter in the neighborhood for declaring its presence periodically, where $p$ is the transmit power which is assumed small. It is worth noting here that the assumption of low-power transmitter is of great importance in most WSN applications.

Consider transmission over only 1 symbol, i.e., $\tau=1$.  Extension to transmission over $\tau>1$ is possible as will be discussed in Sec. \ref{sec:section5}.  Thus, the unquantized signal received by the $s$-th sensor may be written as
\begin{eqnarray}
\mathcal{H}_0: 	y_s = w_s, \hspace{0.5em}
\mathcal{H}_1:	y_s= \sqrt{p} h_s+ w_s
\end{eqnarray}
where $w_s\sim \mathcal{N}(0,1)$ is spatially and temporally independent Gaussian noise,  and $h_s$ is an unknown Rayleigh-fading channel coefficient between sensor $s$ and transmitter, which is assumed constant over one coherence time of channel $T_c>1$. 

After 1-bit quantization, we have 
\begin{align} 
	\nonumber
	\mathcal{H}_0 &: 	  b_s=   \mathcal{Q}_w( {w}_s), \\  
	\mathcal{H}_1 &: 	   b_s= \mathcal{Q}_w(\sqrt{p} h_s + {w}_s)
	\end{align}
where $b_s$ is the binary digit observed at sensor $s$.

\paragraph{\textbf{Signal detection}}
Conditioned on each hypothesis, the binary observations $b_1, b_2,\cdots, b_N $ at different sensors are assumed statistically independent, which is a reasonable assumption( i.e., noise and channels are independent). Accordingly, the detection problem can be expressed as the following binary hypothesis testing:
\begin{align} 
\nonumber
&\mathcal{H}_0:b_1, \cdots,b_n\overset{i.i.d.}{\sim} \operatorname{Ber}(2Q(c)),\\  
&\mathcal{H}_1:b_1, \cdots,b_n\overset{i.i.d.}{\sim} \operatorname{Ber}(2Q  ({c}{(\operatorname{SNR} +1)^{-1}}))
\end{align}
where we have assumed that, under low transmit power $p$, the average received power at each sensor is roughly the same, i.e., $\E {(\sqrt{p} h_1)^2} \approx\E {(\sqrt{p} h_s)^2} \cdots \approx \E {(\sqrt{p} h_n)^2} = \bar{p} \triangleq \operatorname{SNR}$, i.e., approximately same pathloss, which is a sensible assumption in many situations. Note that BDD algorithm relies only on $c$ to compute the test threshold which is fixed and known.

Then each sensor needs to send 1-bit of information $b_s$ to the master node through a multiple access channel (MAC) for the final decision. Assume an error-free channel between sensors and the master node, e.g., a sufficiently good error control coding (ECC) is used\cite{Howard2006}. Then, the master sensor collects all data from all sensors perfectly. Having collected data, the master node thus runs BDD algorithm (Algorithm \ref{alg:alg1}) and reports the decision to the AP (see Fig. \ref{fig:WSN_schematic}).

\section{Numerical Results}
\label{sec:section5}
This section presents some numerical examples to verify the analytical results in this paper. We set $c=1.6$, hence the upper- and lower-window thresholds are $ w_U= 1.6 \sigma_0$ and $w_L=-1.6\sigma_0$.  Consequently, $\theta_0 = 2Q(1.6) \approx 0.11$ \eqref{eq:theta0}. The choice of $c=1.6$ is justified in the next. For the sake of comparison, we simulate the performance under an unquantized system (unquantized observations), employing log-likelihood test which is reduced to $\mathcal{X}_{v}^2$ test (Chi-square test) \cite{detection95}, where $v$ is the number of degrees of freedom.  The former test has been recently used in \cite{HASSAN2020152945} for pilot attack detection in unquantized massive MIMO system.

According to Theorem \ref{thm:PD_WD}, $P_D$ depends on four parameters: $P_F, n,  \alpha$ and the window comparator parameter $c$. The first two parameters are narurally fixed beforehand, whereas the third is unknown (otherwise the test may not be required). Thus optimizing $P_D$  depends on $c$. Since BDD algorithm detects the change in variance between $\mathcal{H}_0$ and $\mathcal{H}_1$, it is natural to select $c$ such that the detection capability is maximized for a small change in variance, i.e., when  $\alpha = \sigma_0/\sigma_1 \in (0,1)$ is large, which in turn leads automatically to maximizing $P_D$ under larger change of variance.  Because of analytical intractability of $P_D$ \eqref{eq:P_D_true_BDJD}, numerical evaluation is used instead.

\begin{figure}[!ht]
	\centering
	\includegraphics[scale=0.44]{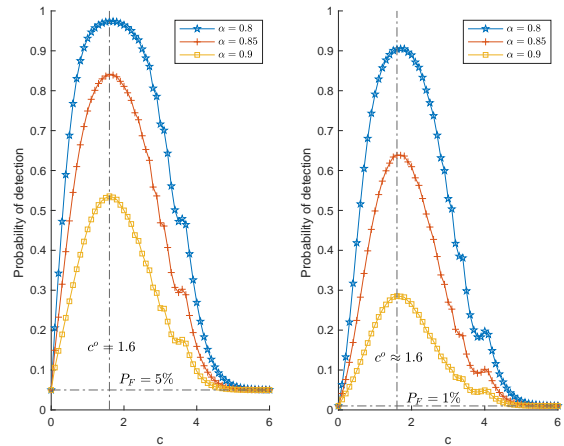}
	\caption{ The optimal $c$ maximizing $P_D$ of BDD algorithm under very small change in variance between $\mathcal{H}_0$ and $\mathcal{H}_1$ when $n=200$. Curves are obtained by Theorem \ref{thm:PD_WD}, where the leftmost plot corresponds to $P_F=5\%$ and the rightmost plot to $P_F = 1\%$.}
	\label{fig:optimal_c} 
\end{figure}

 In Fig. \ref{fig:optimal_c},  we plot $P_D$ against different values of $c$ for different values of $\alpha = 0.8, 0.85,0.9$ (small variance change),  two fixed $P_F =  5\%,  1\%$ and $n=200$. As shown in Fig. \ref{fig:optimal_c},  the optimal $c$  that maximizes $P_D$ is about $1.6$ under both probabilities of false alarm \footnote{ The slight ripple in the tail is an artifact due to  numerical instability.}. It is worth noting that as $c\to 0$ or $c\to \infty$, $P_D$ converges to $P_F$ and hence the detection capability is diminished, verifying the result in Remark \ref{rem:remark1}. Note that this situation is avoidable as $c \in [1,2]$. 
\begin{figure}[!ht]
	\centering
	\includegraphics[scale=0.44]{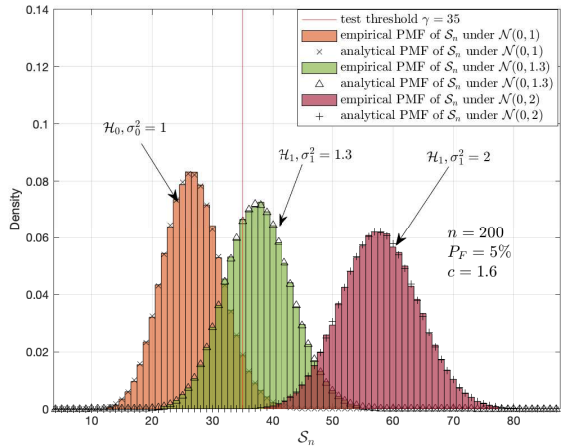}
	\caption{The distribution of test statistic under $\mathcal{H}_0$ and $\mathcal{H}_1$. The analytical results are obtained by \eqref{eq:BinomialPDF}.}
	\label{fig:PMF_Test_Statistics} 
\end{figure}
 
 Fig. \ref{fig:PMF_Test_Statistics} illustrates the distribution of the sum of binary digits (i.e., test statistic $\mathcal{S}_n$) observed at the output of a window comparator in response to two Gaussian processes $y_t \sim \mathcal{N}(0,1)$ ($\mathcal{H}_0$) and \titi{$\tilde{y}_t$} $\sim \mathcal{N}(0,1+\Delta \rm{Var})$ ($\mathcal{H}_1$), where $\Delta \rm{Var}$ represents the change in variance between the two hypotheses and $t=1,2,\cdots,n=200$. As seen, the simulation results match the analytical ones. The results show that the overlapping region of the two distributions shrinks as $\Delta \rm{Var}$ increases, thereby the distinguishability between $\mathcal{H}_0$ and $\mathcal{H}_1$ \titi {is monotonically improving}. This is in line with Corollary \ref{cor:Asymptotic_in_alpha_and_n} where it is shown that $P_D\to 1$ as $\alpha = \sigma_0/\sigma_1=1/\sqrt{1+\Delta \rm{Var}}\to 0$ (i.e., when $\Delta \rm{Var}$ is sufficiently large).
\begin{figure}[!ht]
	\centering
	\includegraphics[scale=0.44]{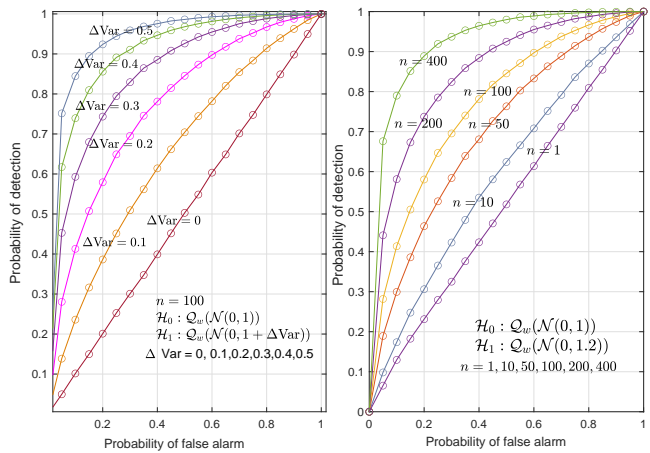}
	\caption{Tradeoff between $P_F$ and $P_D$ for different values of change in variance under two Gaussian processes undergoing 1-bit quantization. Solid lines and markers denote simulated and analytical results (Theorem \ref{thm:PD_WD}), respectively. }
	\label{fig:PF_Vs_PD} 
\end{figure}

Fig. \ref{fig:PF_Vs_PD} shows the tradeoff between $P_D$ and $P_F$ for two Gaussian processes. The performance curves  show that the tradeoff becomes less stringent as the variance of the stochastic process, $\sigma_1^2$, under $\mathcal{H}_1$ and the number of samples, $n$, increase, as shown in Fig. \ref{fig:PF_Vs_PD} (left) and Fig. \ref{fig:PF_Vs_PD} (right), respectively. For instance, Fig. \ref{fig:PF_Vs_PD} (right) suggests that for a fixed change in variance between both hypotheses, the probability of false alarm can be made arbitrarily small while probability of detection converges to 1, asymptotically in the number of observations.  The results are in agreement with the calculations ( i.e., Corollary \ref{cor:Asymptotic_in_alpha_and_n}) where the simulated and analytical results match perfectly. 
\begin{figure}[!ht]
	\centering
	\includegraphics[scale=0.44]{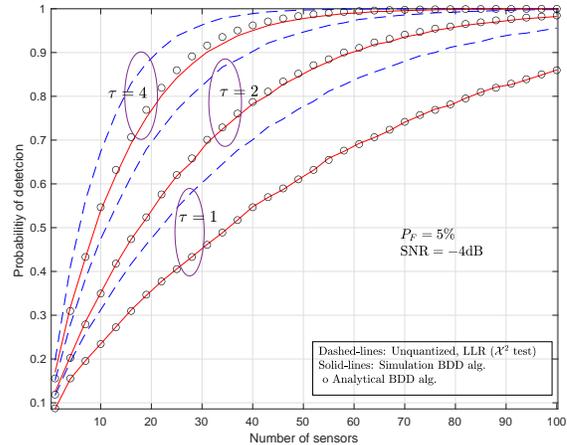}
	\caption{Performance of 1-bit window-comparator based BDD algorithm in WSN for probing a low-power transmitter.}
	\label{fig:WSN_1} 
\end{figure}

In Fig. \ref{fig:WSN_1} we consider a WSN (complex-valued channel model). We show the performance of BDD algorithm run on a master sensor upon collecting data (error-free channel) from other sensors to detect the existence of a low-power transmitter. The transmitter sends a length-$\tau$ constant signal $\vv{x}$  at $\operatorname{SNR} = -4$dB. We equip all sensors (except the master sensor) with window comparators for counting the number of 1’s observed over $\tau =1,2,4$ symbol intervals and reporting the result to the master sensor for a final decision, after which the AP is acknowledged of the decision. Also, the performance under the unquantized system is shown.

\titi{Referring to} Fig. \ref{fig:WSN_1}, the results show that the performance of \titi{the proposed 1-bit window comparator combined with} BDD algorithm is effective regardless its simplicity. With $n=20$ sensors, by letting the transmitter send 4 symbols ($\vv{x} = \frac{1}{\sqrt{2}} [1+j, 1+j, 1+j, 1+j]^{\tran}$) instead of 1 symbol, $P_D$ can be increased from $35\%$ to $80\%$. As a result, increasing the number of observations interval can relax the need for a larger number of sensors to a achieve a satisfactory $P_D$. Compared with an unquantized system \titi{where all sensors are equipped with sufficiently high-resolution ADCs}, the gap in performance decreases gradually as more symbols are transmitted by the transmitter to declare itself. This is because when $\tau>1$, symbols received under $\mathcal{H}_1$ exhibit some correlation, incurring a loss of performance for both systems and hence narrowing down the gap. 

It is worth mentioning that the case $\tau=1$ (one symbol) gives rise to uncorrelated samples in space and time, hence $P_D$ in Theorem \ref{thm:PD_WD} matches perfectly with the simulation result. When $\tau>1$, there is some temporal correlation since we assume channel constant during this short-length transmission. Therefore, the analytical $P_D$ which is derived under uncorrelated samples deviates slightly from the simulation results, though it is not noticeable \footnote{When transmit power is sufficiently small, correlation between samples can be negligible and hence the observed bits can be well approximated as i.i.d. Bernoulli random variables.}.  The analytical probability of detection thus obtained is sufficiently accurate even under correlated observations. 

\begin{figure}[!ht]
	\centering
	\includegraphics[scale=0.44]{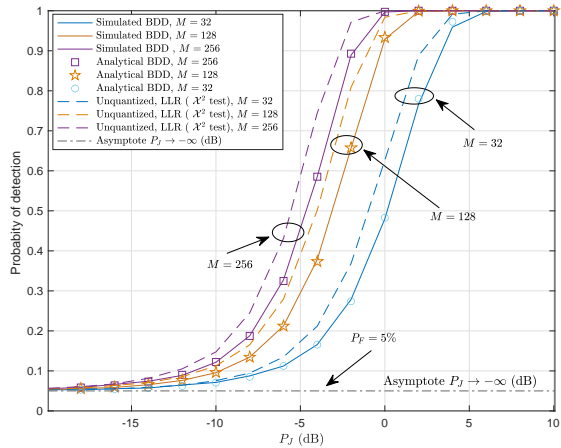}
	\caption{Performance of 1-bit window-comparator based BDD algorithm in massive MIMO system for detecting a single-antenna jammer. $N=1$ (one coherence block), $K=\tau=5$ $\beta_i=1, p_i=0$ dB.}
	\label{fig:BDD_MassiveMIMO_re1} 
\end{figure}

Fig. \ref{fig:BDD_MassiveMIMO_re1} shows the performance of jamming detection in a massive MIMO system. We assume $K=5$ users, $\tau=K$ (pilot length) and users use mutually orthogonal pilots where all pilot symbols have unit modulus. For simplicity, we assume $\beta_J=\beta_1=\cdots=\beta_K=1$ and $p_1= \cdots=p_K=0$dB. Only one coherence channel block ($N=1$) is considered, and all pilot symbols are exploited for detection. Conditioned on $\mathcal{H}_0$, the case $K = \tau$ renders all received (unquantized) pilot symbols i.i.d. Gaussian random variables having zero-mean and common variance $\sigma_0^2 = Kp+1$ and hence the corresponding binary digits are i.i.d. Bernoulli random variables.  
\begin{table}[!ht]
	\caption{Simulation parameters of Fig. \ref{fig:BDD_MassiveMIMO_re1} for BDD algorithm (1-bit quantized system) and  $\mathcal{X}_{2 \tau M}^2$ test (unquantized system) with $P_F=5\%$.}
	\centering
	\begin{tabular}{ccccc}
		\hline
		$M$& $\gamma$ & $\zeta$ & $\gamma^{\mbox{unq.}}$($\times 10^3$)&\\
		\hline
		32&  44&  0.026& 1.0882& \\
		128&  159&  0.5989 & 4.093&  \\
	256	&  307& 0.7024 &  8.0365& \\
	\hline
	\end{tabular}
\label{tab:table1}
\end{table}

For comparison, we also simulate the corresponding unquantized system. The parameters of BDD algorithm (for 1-bit quantized system) and $\mathcal{X}_v^2$ test (for unquantized system) are shown in Table \ref{tab:table1}.  It is clear that the simulation results match perfectly with the analytical ones (Theorem \ref{thm:PD_WD}).  Interestingly, we observe that our proposed \titi{1-bit window comparator combined with BDD algorithm} can achieve a significant fraction of the performance achieved when the BS has direct access to unquantized samples. As seen for most cases of interest (say, a region below 0dB), the performance gap is about $10 \% $. This verifies that our proposed 1-bit window comparator combined with BDD algorithm is efficient for jamming detection.

\titi{Referring to Table \ref{tab:table1}, it is interesting to note that the randomization parameter $\zeta$ tends to 0 or 1 as the number of BS antennas, $M$, decreases or increases, respectively. In general, this is the case when the number of observations, $n=2MN$, decreases or increases, respectively. According to Algorithm \ref{alg:alg1}, this implies that the decision process at the BS becomes nonrandom when $n$ is either too small or too large. For instance, when $M$ is sufficiently large, $\mathcal{H}_1$ is favored over $\mathcal{H}_0$ most often. The reason for this is that, as $n=2MN$ increases, the distribution of the sufficient statistic $\mathcal{S}_n$ gets closer and closer to the distribution of a continuous random variable, i.e., see discussion in Subsec. \ref{subsec:Analysis of BDD algorithm}.}

\begin{figure}[!ht]
	\centering
	\includegraphics[scale=0.44]{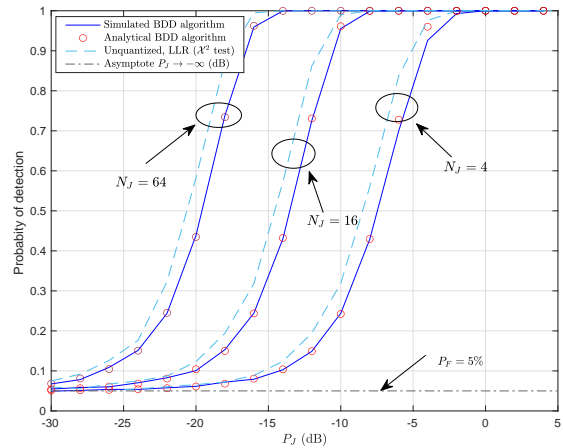}
	\caption{Performance of 1-bit window-comparator based BDD algorithm in massive MIMO system for detecting a multiple-antenna jammer.  $N=1$ (one coherence block), $K=\tau=5$ $\beta_i=1, p_i=0$ dB.}
	\label{fig:BDD_MassiveMIMO_res2} 
\end{figure}

In Fig. \ref{fig:BDD_MassiveMIMO_res2} we consider the same scenario as in Fig. \ref{fig:BDD_MassiveMIMO_re1} except the jammer uses $N_J$ antennas. This can also capture a situation in which there are multiple jammers equipped with multiple antennas, such that the total number of antennas of all jammers is $N_J$. We also assume that the jammer sends a random pilot signal from its antennas (which is not a strict condition). As expected, $P_D$ improves gradually as $N_J$ increases since the performance of our detection mechanism improves when change in variance between $\mathcal{H}_0$ and $\mathcal{H}_1$ increases (Corollary \ref{cor:Asymptotic_in_alpha_and_n}). 

\section{Conclusion}
\label{sec:section6}

\titi{In this work, a binary hypothesis detection mechanism, comprising 1-bit window comparator and bit density detection (BDD) algorithm, has been proposed  for detecting the presence or absence of an unknown signal in a 1-bit quantized MIMO receiver. The incoming signal is quantized by an adapted and well-designed 1-bit window comparator, generating a well-structured Bernoulli process with a \emph{fixed} success rate under the model in null hypothesis ($\mathcal{H}_0$). Thus, our BDD algorithm exploits this knowledge to detect any deviation from the default success rate, where this deviation is highly likely to be induced by a model in the alternative hypothesis ($\mathcal{H}_1$). We also gave the probability of detection of our algorithm in Theorem \ref{thm:PD_WD}.}

We showed that the simulated and analytical results match correctly. \titi{This paper has clearly shown that the proposed 1-bit window comparator combined with BDD algorithm is very efficient in detecting jamming attack in 1-bit quantized massive MIMO system and probing the presence of a low-power transmitter in a wireless sensor network. In both cases, it was shown that our simple receiver relying on a coarsely quantized signal can achieve a high percentage of the performance achieved with no quantization. In general, the findings are of direct practical relevance to binary hypothesis testing relying on binary measurements at a receiver.}

In this research, we only considered the uncorrelated Bernoulli process under simple hypotheses because of its tractability. \titi{In a future research, we intend to concentrate on the general correlated process}, which relaxes the constraint on the deliberate signal design and exploits correlation for further performance improvement.

\bibliographystyle{IEEEtran}

\bibliography{main}

\end{document}